\documentstyle[aps,multicol,epsf]{revtex}
\newcommand{\be}{\begin{equation}}
\newcommand{\ee}{\end{equation}}
\begin{document}
\preprint{SNUTP }
\draft
\title{Anomalous Height Fluctuation Width in Crossover \\
from Random to Coherent Surface Growths \\} 
\author {K.~Park and B.~Kahng \\}
\address {Department of Physics and Center 
for Advanced Materials and Devices, \\ 
Kon-Kuk University, Seoul 143-701, Korea \\} 

\maketitle
\thispagestyle{empty}

\begin{abstract}
We study an anomalous behavior of the height 
fluctuation width in the crossover from random 
to coherent growths of surface for a stochastic  
model. In the model, random numbers are 
assigned on perimeter sites of surface, 
representing pinning strengths of disordered media. 
At each time, surface is advanced at the site having 
minimum pinning strength in a random subset of system 
rather than having global minimum.  
The subset is composed of a randomly 
selected site and its $(\ell-1)$ neighbors. 
The height fluctuation width $W^2(L;\ell)$ exhibits 
the non-monotonic behavior with $\ell$ and it has  
a minimum at $\ell^*$. 
It is found numerically that $\ell^*$ 
scales as $\ell^*\sim L^{0.59}$, and the height fluctuation width 
at that minimum, $W^2(L;\ell^*)$, scales  
as $\sim L^{0.85}$ in 1+1 dimensions. 
It is found that the subset-size $\ell^*(L)$ is 
the characteristic size of the crossover from 
the random surface growth in the KPZ universality, 
to the coherent surface growth in the directed percolation 
universality.   
\end{abstract} 

          
\begin{multicols}{2}
\narrowtext 
Recently subjects in the field of non-equilibrium surface
growth have been attractive, which is due to the 
interest in theoretical classification of universality 
and also due to their applications to various physical 
phenomena such as crystal growth, molecular beam epitaxy, 
vapor deposition, and biological evolution, $etc$ \cite{review}.
The interesting feature of non-equilibrium surface growth is  
the nontrivial scaling behavior of the height fluctuation width 
\cite{family}, 
i.e.,
\be
W^2(L,t)= \langle {1 \over L^{d'}} \sum_{i} (h_i - \bar{h} )^2 
\rangle  \sim L^{2\alpha} f(t/L^z),
\ee
where $h_i$ is the height of site $i$ on substrate. 
Here, $\bar h$,~$L$, and $d'$ denote mean height, 
system size, and substrate dimension, respectively. 
The symbol $\langle \cdots \rangle$ stands for statistical average.
The scaling function behaves as $f(x) \rightarrow \hbox{constant}$
for $x \gg 1$, and $f(x) \sim x^{2\beta}$ for $x \ll 1$ with 
$\beta=\alpha /z$.
The exponents, $\alpha$, $\beta$ and $z$ are called 
the roughness exponent, the growth exponent,
and the dynamic exponent, respectively. \\
                              
The Kardar-Parisi-Zhang (KPZ) equation \cite{kpz} 
was introduced to account for the effect of sideway 
growth, which is written as  
\be
{{\partial h(x,t)} \over {\partial t}}=\nu \nabla^2 h(x,t)
+{{\lambda \over 2}} (\nabla h(x,t))^2+\eta(x,t),
\ee
where $\eta(x,t)$, called thermal noise, is assumed as 
white noise, 
$\langle \eta(x,t) \rangle =0$ and 
$\langle \eta(x,t) \eta(x',t') \rangle =2D \delta^{d'}(x-x') \delta(t-t')$
with noise strength $D$. 
Many stochastic models in the KPZ universality class have been 
introduced \cite{review}. Among them, the  
restricted solid-on-solid (RSOS) model was introduced 
by Kim and Kosterlitz \cite{rsos}, satisfying the scaling 
relation, $\alpha_t+z_t=2$. The subscript means that the exponents 
are for $thermal$ noise.  
In the RSOS model, a particle is deposited at a randomly 
selected site as long as the height difference $\Delta h$ between nearest 
neighbor columns remains as $\Delta h \le 1$ even after deposition, 
otherwise, the particle is excluded from deposition.
On the other hand, one may modify the dynamic rule of the 
RSOS model by replacing the exclusion by avalanche: 
At each time step, a particle is deposited at a randomly 
selected site, and avalanche may occur successively 
to nearest neighbor sites as long as the height difference between 
nearest neighbors, $\Delta h > 1$. 
Then it was argued that this model also belongs 
to the KPZ universality class \cite{sne}, but the model requires 
relatively large system size to see its asymptotic behavior. 
Let us call the former model the RSOS-A model, 
and the latter model the RSOS-B model. The surface-growth in the 
KPZ universality class is called the random surface-growth.\\ 

Physical properties of growing surface in disordered media 
are different from those of the thermal KPZ equation, Eq.~(2). 
In order to account for the effect of disorder in porous media, 
quenched noise\cite{levine}, which depends on position $x$ and 
height $h$, replaces thermal noise in Eq.~(2).   
Then the quenched KPZ (QKPZ) equation is written as 
\be
{{\partial h(x,t)} \over {\partial t}}=\nu \nabla^2 h(x,t)
+{{\lambda \over 2}} (\nabla h(x,t))^2+\eta(x,h),  
\ee
where the noise satisfies, $\langle \eta(x,h) \rangle =0$, and   
$
\langle \eta(x,h) \eta(x',h') \rangle 
=2D \delta^{d'}(x-x') \delta (h-h').
$
Stochastic models associated with the QKPZ 
equation have been introduced \cite{boston,tang}. 
The models show that the surface of the QKPZ equation 
in 1+1 dimensions belongs to the directed percolation (DP) 
universality. The roughness exponent $\alpha_q$ in 
the QKPZ equation is given by the ratio of the correlation 
length exponents in perpendicular and parallel directions, 
$\nu_{\perp}$ and $\nu_{\parallel}$, of directed percolating clusters, 
which is $\alpha_q=\nu_{\perp}/ \nu_{\parallel}\approx 0.63$.  
On the other hand, recently Sneppen introduced a stochastic 
model in which surface grows coherently \cite{sne}. 
In that model, random numbers, representing disorder of porous media, 
are assigned on each perimeter site of surface. 
Surface is advanced at the site having global minimum among 
the random numbers. The avalanche rule is then applied 
successively to nearest neighbor sites as long as 
$\Delta h > 1$. Random numbers at the columns 
with increased heights are updated by new ones.  
The Sneppen model also belongs to the DP universality, and 
the roughness exponent is $\alpha_s \approx 0.63$
in 1+1 dimensions\cite{grassberger}. The surface-growth  
of the Sneppen model is called the coherent surface-growth. \\ 

The coherent surface-growth model is closely related to 
the self-organized evolution model for biological systems \cite{evo}. 
In the evolution model, one considers random numbers  
assigned in one-dimensional array, which represent 
fitness of each species. Mutation of species is described 
in the model by updating random numbers.  
The updating occurs at the site having global minimum random number 
and its two nearest neighbors, and the random numbers at those  
sites are replaced by new ones at each time.  
Then as times go on, relatively small random 
numbers disappear by the updating process, 
and the distribution of random numbers exhibits a self-organized 
critical behavior. On the other hand, one may think 
of the situation where biological evolution 
is not motivated in globally optimized manner, but 
it may be driven by optimization within 
$finite$ region out of entire system. 
This is analogous to the case that spin glass system is in 
meta-stable state within finite relaxation time rather than in 
globally stable state. Motivated by this idea, 
in this paper, we introduce a surface growth 
model in disordered media, where surface growth occurs at 
the site having the minimum random number in a $subset$ of the entire 
system rather than having the global minimum random number. 
We think this model might be relevant to the case where
the relaxation of surface growth in disordered media is not fast enough 
to spread into the whole system, so that surface growth is driven  
not in globally optimized manner, but in locally optimized manner.\\ 

To be specific, the model we consider in this paper is defined 
as follows: First, we consider one dimensional flat substrate 
with system size $L$. Random numbers are assigned on each site, 
which represent energy barriers (fitness) in the evolution model 
or pinning forces in the surface growth model by Sneppen. 
Second, we select a site randomly, and 
consider the subset composed of $\ell\equiv 2r+1$ elements, 
the randomly selected site and its $2r$ neighbor sites 
within distance $r$. It is worthwhile to note that the subset 
is regarded as a random sample because the site in the middle 
of the subset was selected at random. The subset is formed 
instantaneously, and its territory might overlap with 
subsequent one as shown in Fig.~1.  
Next, surface is advanced at the site having minimum random 
number among the $\ell$ elements in the subset, and 
the avalanche process is followed successively at its neighbor sites 
to keep the RSOS condition, $\Delta h \le 1$, and it may 
spread out over the boundary of the subset.  
Finally, the random numbers at the sites with increased height 
are updated with new ones. The dynamic rule of the model is 
depicted in Fig.~1. 
When $\ell=1$, this model corresponds to the RSOS-B model 
in the KPZ universality, whereas when $\ell=L$, it does to the 
Sneppen-B model in the DP universality. 
Accordingly one may see the crossover behavior from the KPZ 
limit to the DP limit as increasing $\ell$. \\ 

Since the roughness exponent $\alpha_t=1/2$ in the KPZ limit is 
smaller than the one $\alpha_q \approx 0.63$ in the DP limit, 
one may expect at a glance that the height fluctuation 
width $W^2(L;\ell)$ in steady state increases monotonically 
with increasing $\ell$. 
However, we found the anomalous behavior numerically 
that $W^2(L;\ell)$ decreases with increasing $\ell$ 
for small $\ell$, and increases for large $\ell$ 
as shown in Fig.~2. The minimum of $W^2(L;\ell)$ becomes steeper 
and its location, $\ell^*/L$, which was rescaled by system size $L$, 
approaches to zero as $L$ increases. 
It is found that the location of the minimum scales 
as $\ell^* \sim L^{0.59}$, and the height fluctuation 
width at this minimum scales as $W^2(L;\ell^*) \sim L^{0.85}$, 
which are shown in Figs.~3 and 4.  
The estimated values of $\ell^*$ and $W^2(L;\ell^*)$  
for different system sizes are tabulated below.\\ 

Table 1. Numerical values of the location $\ell^*$ and the height 
fluctuation width $W^2(L;\ell^*)$ at the minimum 
for different system sizes. 
                                                                          
\begin{center}
\begin{tabular}{|c|c|c|c|c|c|}
\hline
 $L$ & $\ell^*$ & $W^2(L;\ell^*)$ \\
\hline
64 & 11 & 5.9  \\
\hline
128 & 19 & 11.2  \\
\hline
256 & 27 & 20.3  \\
\hline
512 & 39 & 35.8  \\
\hline
1024 & 59 & 62.5 \\
\hline
2048 & 95 & 110.0 \\
\hline
\end{tabular}
\end{center}
\smallskip

The anomalous behavior may be attributed to the two effects, 
the random effect for small $\ell$ and the coherent effect for 
large $\ell$. For small $\ell$, surface grows by random deposition with 
avalanches, and it belongs to the KPZ universality.  
Thus the height fluctuation width depends on system size as 
$W^2(L;\ell) \sim L^{2\alpha_t}$ with $2\alpha_t=1$, 
however, it would also depend on subset 
size $\ell$. In order to find out the $\ell$-dependent behavior 
of $W^2(L;\ell)$ in phenomenological level,  
we plot $W^2(L;\ell)$ versus $\ell$ in double logarithmic 
scales for several values of $L$ as shown in Fig.~5. 
The slopes are measured to be $\approx -0.2$ 
for large system sizes $L=1024$ and $2048$. 
Based on this measurement, $W^2(L;\ell)$ is written as 
$W^2(L;\ell) \sim \ell^{-0.2}L$ for small $\ell$. 
This result is contrary to the one, based on the 
coarse-graining scaling argument, $W^2(L;\ell) 
\sim (L/\ell)^{2\alpha_t} \ell^{2\alpha_q}$, which exhibits the 
increasing behavior of $W^2$ with increasing $\ell$.  
On the other hand, when $\ell$ is large enough, surface updating 
is initiated mainly at the site having global minimum random number  
of entire system. When the site of global minimum 
is selected, which occurs with probability $\ell /L$, 
the surface becomes correlated by the Sneppen dynamics, 
however, when the site of the global minimum 
is not selected, the correlation formed by the Sneppen's dynamics 
is relaxed.  
Thus, when $\ell$ is large enough so that the contribution by 
the Sneppen dynamics is sufficiently dominant, one may 
write the dominant term of the $averaged$ height fluctuation width as 
$W^2(L;\ell) \sim ({\ell \over L}) L^{2\alpha_q}$. 
That is because the statistical average was taken over the quantity
of the square of mean-height deviation in Eq.~(1).  
Combining the two asymptotic behaviors obtained 
in phenomenological level, 
the height fluctuation width is written as 
\be
W^2(L;\ell) \sim {\ell^{-0.2}} L + \big({\ell \over L}\big) L^{1.26}.
\ee    
The two terms exhibit the competing behavior with 
respect to $\ell$, which yields the anomalous behavior. 
Taking the derivative with respect to 
$\ell$, the location $\ell^*$ of the minimum 
is obtained as that $\ell^* \sim L^{0.62}$, which is close to  
the numerical measurement, $\ell^* \sim L^{0.59}$. 
We also examine the size-dependent behavior of 
the height fluctuation width at the 
minimum by plugging $\ell^*$ into Eq.~(4), and 
obtain that $W^2(L;\ell^*) \sim L^{0.88}$. This result is also 
close to the numerical measurement, $W^2(L;\ell^*) \sim L^{0.85}$.
The numerical estimations for $\ell^*$ and $W^2(L;\ell^*)$ are 
better explained by minimizing the formula, 
\be
W^2(L;\ell) \sim \big({{\ell^{2\alpha_t}} \over {\ell^{2\alpha_q}}}\big) 
L^{2\alpha_t}+\big({\ell \over L}\big) L^{2\alpha_q},
\ee    
however, the derivation of this formula is not clear. 
\\
 
In order to understand physical meaning of the characteristic 
subset-size $\ell^*$, we plot $W^2(L;\ell)$ versus $L$ up to $L=2048$ 
in double logarithmic scales for typical subset sizes, $\ell=5$ 
and 41 in Fig.~6. 
The size of $\ell=5$ corresponds to the case where it is smaller 
than $\ell^*(L)$ for all system sizes $L$ used in Fig.~6. 
However, the size of $\ell=41$ does to the case where it is smaller 
than $\ell^*(L)$ in part for $L=1024$ and $2048$, 
close to $\ell^*(L)$ for $L=512$, but larger than 
$\ell^*(L)$ in part for $L=64$ and $128$. 
For the case of $\ell=5$, all data are on a straight line, 
whereas, for the case of $\ell=41$, 
forming a straight line breaks 
down for relatively smaller system sizes $L=64$ and $128$.   
Fig.~6 suggests that $\ell^*(L)$ be the characteristic 
subset-size such that when $\ell < \ell^*(L)$, 
the roughness of overall surface is determined by 
random effect, whereas when $\ell > \ell^*(L)$, it is done by  
coherent effect. Accordingly, the subset size $\ell^*$ 
has the meanings of not only the location of the minimum of 
the anomalous height fluctuation width, but also the critical size 
at which the crossover from random to coherent surface-growths occurs. 
Also note that even for the case of $\ell > \ell^*(L)$, 
the surface height fluctuation width does not behave as
$W^2(L;\ell) \sim L^{2\alpha_q}$. The roughness 
exponent has smaller value than $2\alpha_q$ as appeared 
in Fig.~6, and the numerical estimation of the roughness exponent 
is likely to be ${2\alpha_q-1}$ as Eq.~(4) for fixed $\ell$.  
The behavior of $W^2(L;\ell) \sim L^{2\alpha_q}$ occurs 
when $\ell$ also increases as $L$ increases.      
In thermodynamic limit, $L \rightarrow \infty$,   
the characteristic size $\ell^*$ goes to infinity, so that 
for finite $\ell$, the roughness of overall surface is 
determined by random effect, 
and the surface is described in the KPZ universality.  
Next, we examine the height-height correlation function, 
\be 
C(r;L,\ell)\equiv \langle {1 \over L} \sum_x (h(x+r)-h(x))^2 \rangle,  
\ee
which is defined for fixed $L$, and $\ell$.  
In Fig.~7, we plot $C(r)/l^{2\alpha_q}$ versus $(r/\ell)$ in 
double logarithmic scales for several values of $\ell$, and $L=2048$. 
In Fig.~7, the data are well collapsed for $r/\ell < 1$, 
while they are not collapsed for $r/\ell > 1$.    
Accordingly, the coherent surface growth occurs within 
the range of $r < \ell$, however, the roughness of overall surface is 
determined by the criterion depending on $\ell^*(L)$ and $L$ 
above. \\  
 
It would be interesting to study the $\ell$-dependent 
behavior of dynamic properties of the height fluctuation width 
$W^2(L,t;\ell)$. The study is based on numerical simulations 
for fixed system size, say, $L=1024$. As shown in inset of Fig.~8, 
there exist four distinct regimes for $W^2(L,t;\ell)$.  
In the first regime, $W^2(L,t;\ell)$ increases according to the 
Poisson distribution, and $W^2(L,t;\ell) \sim t^{2\beta_1}$ 
with $2\beta_1=1$. The first regime terminates at $t_1$, 
which is independent of subset-size $\ell$.  
In the second regime, surface becomes correlated 
by the coherent effect, which is caused by the selection of minimum 
random number in selected subset, 
however, the decorrelation also occurs simultaneously 
by the random effect, which is caused by the selection of the random subset. 
Since the value of the dynamic exponent $z_s=0.63$ for the Sneppen 
dynamics is smaller than the one $z_t=1.5$ for the KPZ dynamics, 
the coherent effect spreads faster than the random effect 
in early times. 
Thus the growth exponent $2\beta_2$ in the second regime 
has the value more likely close 
to the Sneppen value, $2\beta_s=2$, however, the value is 
a little bit smaller by the decorrelation by the random effect.  
The growth exponent $2\beta_2$ 
depends on subset-size $\ell$ as tabulated in Table 2. 
Based on the measurement in Fig.~9, the growth exponent is likely to 
depend on $\ell$ as $2\beta_2 \sim (0.1)\log \ell$ for $\ell < \ell^*$, 
however for $\ell > \ell^*$, the value of the growth exponent is 
expected to be close to the one of the Sneppen dynamics as shown in 
Fig.~8. 
The second regime terminates at $t_2$.  In Fig.~10, 
the threshold times are likely to scale as $t_2 \sim \ell^{0.37}$ 
for $\ell < \ell^*$. The values of the height fluctuation width at 
the threshold value $t_2$ are likely to scale 
as $W_2^2(L,t_2; \ell) \sim \ell^{0.92}$ as shown in Fig.~11.
In the third regime, the random effect appears much dominantly, and 
the coherence of surface formed in the second regime becomes 
decorrelated in this regime. 
As subset size is smaller, the third regime is much 
dominant, and the growth exponent $\beta_3$ becomes much closer 
to the KPZ value, whereas as subset size is larger, 
the decorrelation effect becomes much weaker, so that the growth 
exponent $\beta_3$ becomes smaller. 
It is likely that $2\beta_3 \sim (-0.37)\log \ell$ as shown in Fig.~12. 
The third regime terminates at $t_3$. The numerical values of $t_3$ 
for different sizes of $\ell$ locate too closely for small $\ell$ 
to be measured numerically.\\ 

Table 2. Numerical estimation for the values of the growth exponents, 
the threshold times, and the height fluctuation widths 
for various subset sizes $\ell$.\\ 

\begin{center}
\begin{tabular}{|c|c|c|c|c|c|}
\hline
$\ell$ & $2\beta_2$ & $\log_{10}(t_2)$ & $\log_{10}(W_2^2)$ & $2\beta_3$ \\ 
\hline
3 & 1.12 & -0.10 & 0.20 & 0.66   \\
\hline
5 & 1.16 & -0.10 & 0.20 & 0.64  \\
\hline
9 & 1.23 & 0.00 & 0.33 & 0.60  \\
\hline
17& 1.31 & 0.10 & 0.55 & 0.50   \\
\hline
33 & 1.40 & 0.20 & 0.80 & 0.38   \\
\hline
65 & 1.59 & 0.31 & 1.12 & 0.28  \\
\hline
129 & 1.68 & 0.42 & 1.60 & 0.17  \\
\hline
257 & 1.78 & 0.60 & 1.80 & 0.17  \\
\hline
\end{tabular}
\end{center}
\smallskip

We also investigated the $\ell$-dependent behavior of 
the distribution of random numbers after reaching 
saturated state.  
As shown in Fig.~13, the distribution is flat for $\ell=1$, and 
exhibits a critical behavior for $\ell=L$. Between the two limits, 
the distributions look like a rounded step function. 
It would be interesting to note that all distribution 
functions for different $\ell$ pass through a specific value 
of random number $B_c$, which corresponds to the threshold of  
the self-organized critical state \cite{grassberger}. 
The value of $B_c$ equals to $1-P_c=0.462$, 
where $P_c$ is the directed percolation threshold. \\ 

Recently, the crossover behavior from the random surface growth 
to the coherent surface growth was considered by Vergeles \cite{temp}. 
In that study, surface growth occurs at site $x$ on 
substrate with the probability, $P(x) \sim e^{-q(x)/T}$, 
where $q(x)$ is a random pinning strength of site $x$, 
and is also updated with height advance. $T$ is temperature. 
It was found that the surface of the model reduces to 
the one of the Sneppen dynamics when $T=0$, however, 
for $T \ne 0$, it does to the one of the KPZ universality. 
The temperature $T$ plays a role of tuning parameter for the 
crossover behavior. However, since the tuning parameter is 
in the form of exponential function, it is hard to see finite 
size dependent behavior of the crossover which is  
very sensitive to tuning the parameter as we studied in 
this paper. 
Nevertheless, the anomalous behavior may be observed barely 
in the plot of $W(L;T)$ versus $L$ for different temperatures, 
Fig. 1(b) in Ref.~11, where the curves of $W(L,T)$ cross to each other, 
However the crossover behavior has not been remarked in Ref.~11.\\ 

In summary, we have introduced a stochastic model 
for surface growth, which is a generalization of 
the restricted solid-on-solid model in the KPZ universality 
and the Sneppen model in the directed percolation limit, 
and have investigated the crossover of the two limits. 
Deposition occurs at the site having minimum of random 
numbers within finite subset rather than of entire system. 
The subset is composed of $\ell$ elements, a randomly selected site 
and its $\ell-1$ neighbors. 
Changing subset-size $\ell$, 
the height fluctuation width exhibits the anomalous behavior 
having a minimum. 
The anomalous behavior is due to the two competing effects, 
the random effect for small $\ell$ and the coherent effect 
for large $\ell$. The minimum of the surface height fluctuation width,  
locating at $\ell^* \sim L^{0.59}$, is scaled as 
$W^2(L;\ell^*) \sim L^{0.85}$ in 1+1 dimensions. 
The characteristic subset size $\ell^*(L)$ has 
the meaning that for $\ell < \ell^*(L)$, the surface grows 
randomly and belongs to the KPZ universality, 
whereas for the opposite case, surface grows coherently.
The dynamic properties of the crossover have also been investigated. 
In early stage of growth, surface becomes correlated according 
to the Sneppen dynamics, and in late stage, the surface 
correlation is relaxed by the random process. 
The phenomenon of the dynamic correlation-decorrelation behavior 
also appears in a stochastic model \cite{kahng} 
for the flux line dynamics with transversal and longitudinal fluctuations,  
which might be described by the coupled quenched KPZ 
equation \cite{flux}. \\ 

One of the authors (B.K.) thanks Drs. A.-L. Barab\'asi 
and D. Dhar for helpful discussions. 
This work is supported in part by the NON DIRECTED RESEARCH FUND, 
the Korea Research Foundation, in part by the BSRI program 
of Ministry of Education, Korea, and in part by the KOSEF 
through the SRC program of SNU-CTP.  \\


\begin{figure}
\centerline{\epsfxsize=7cm \epsfbox{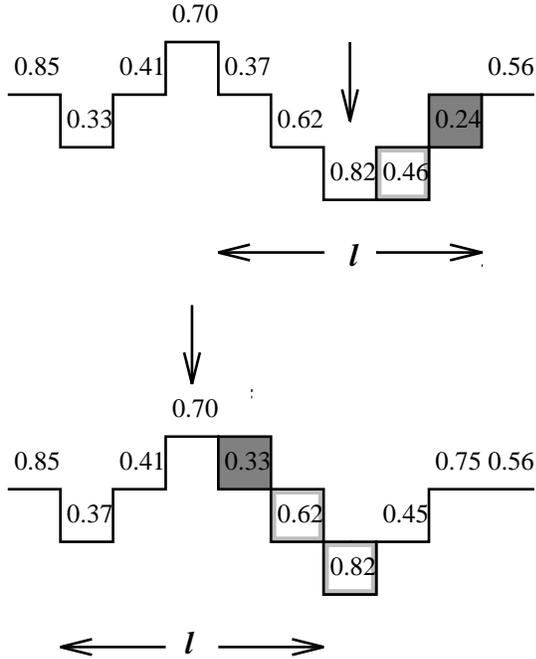}}
\vspace{.5cm}
\caption{
Schematic representation of the stochastic rule. 
The sites arrowed are randomly selected sites.
The dark squares denote the sites of minimum random number 
within subset sized $\ell=5$. The white squares denote the 
sites updated by avalanches.
The subset could overlap with the one (lower one) 
of subsequent time.}
\label{fig1}
\end{figure}
\begin{figure}
\centerline{\epsfxsize=8cm \epsfbox{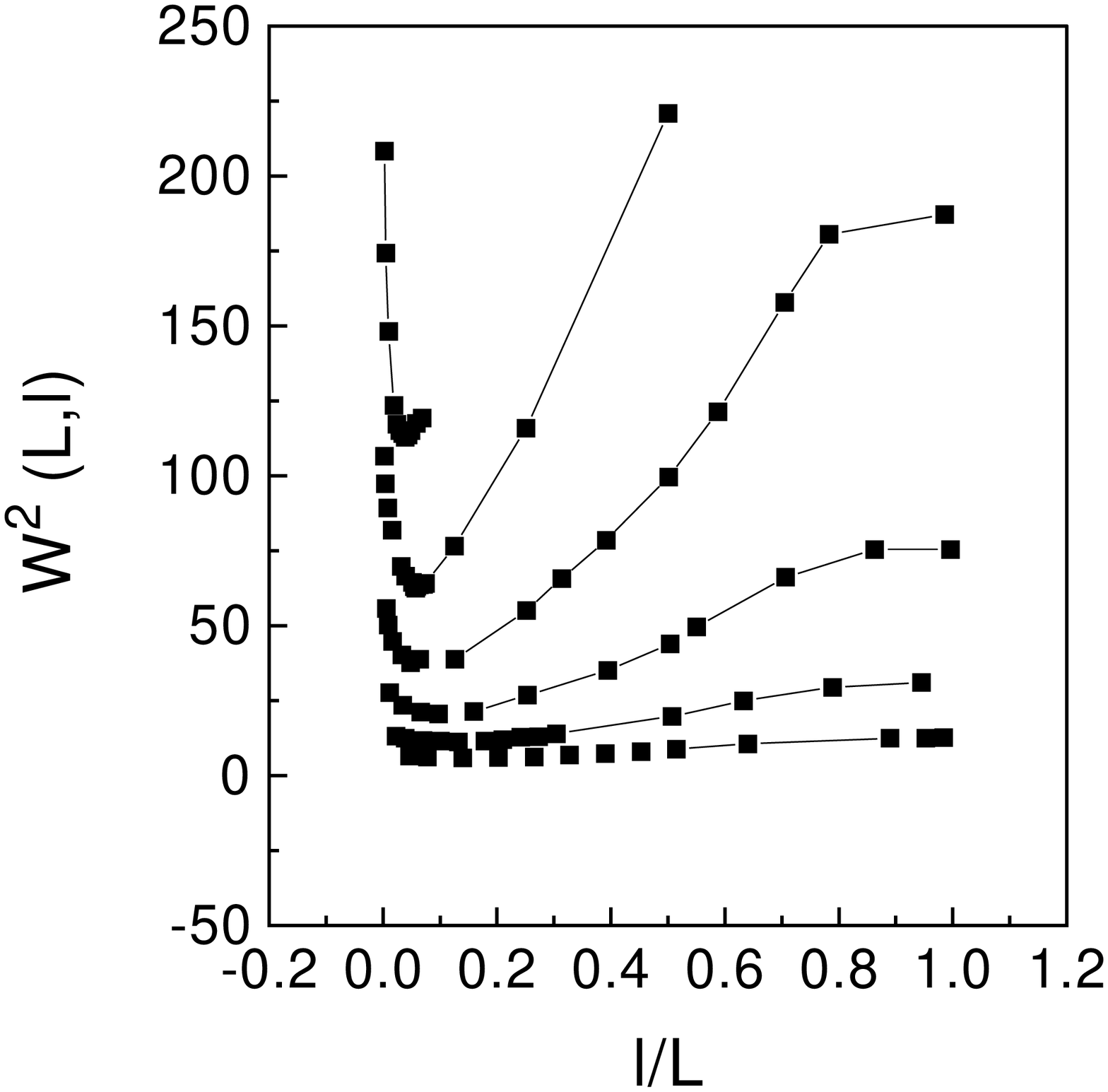}}
\caption{ Plot of the surface height fluctuation 
widths $W^2(L;\ell)$ versus subset size $\ell /L$ rescaled 
by system size $L$ in steady state. 
Numerical data are for system sizes $L = 64, 128, 256, 512, 1024$ and 
$2048$ from the bottom.}
\label{fig2}
\end{figure}
\begin{figure}
\centerline{\epsfxsize=8cm \epsfbox{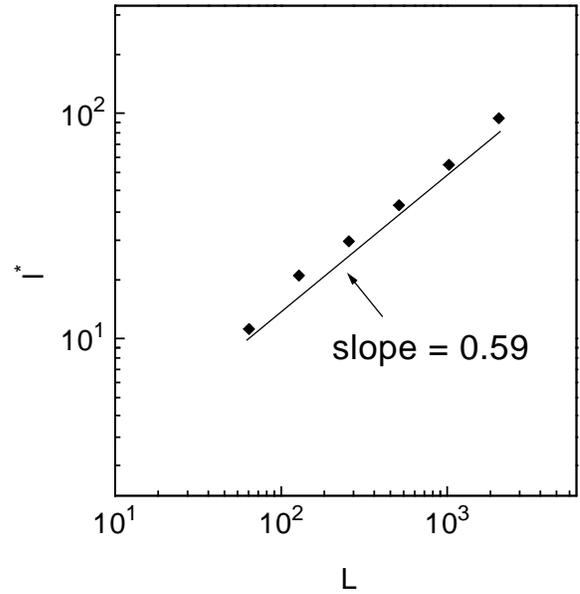}}
\caption{Double-logarithmic plot of the estimated location $\ell^*$ 
of the minimum versus system size $L$. The data are for 
$L=64, 128, 256, 512, 1024$, and $2048$.}  
\label{fig3}
\end{figure}
\begin{figure}
\smallskip 
\centerline{\epsfxsize=8cm \epsfbox{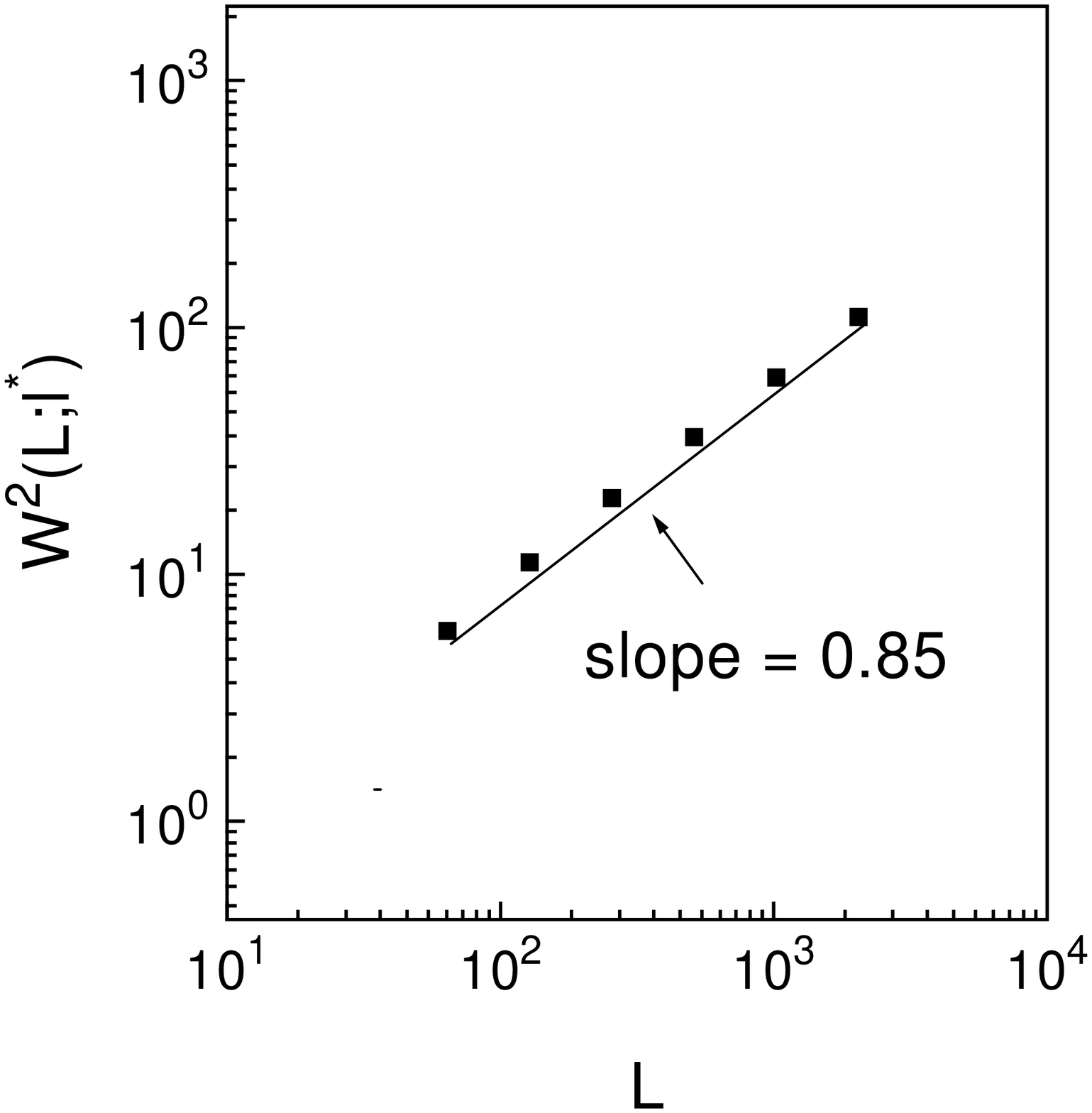}}
\caption{Double-logarithmic plot of the surface height fluctuation 
width at the minimum position $W^2(L;\ell^*)$ versus system size $L$. 
The data are for $L=64, 128, 256, 512, 1024$, and $2048$.} 
\label{fig4}
\end{figure}
\begin{figure}
\smallskip
\centerline{\epsfxsize=8cm \epsfbox{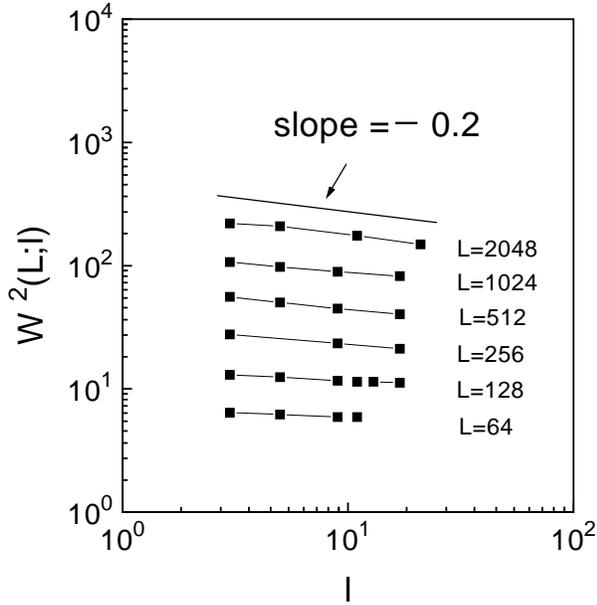}}
\caption{Double-logarithmic plot of $W^2(L;\ell)$ versus $\ell$ 
for various system sizes. The data seem to be on straight 
lines with slope $-0.2$.} 
\label{fig5}
\end{figure}
\begin{figure}
\smallskip
\centerline{\epsfxsize=8cm \epsfbox{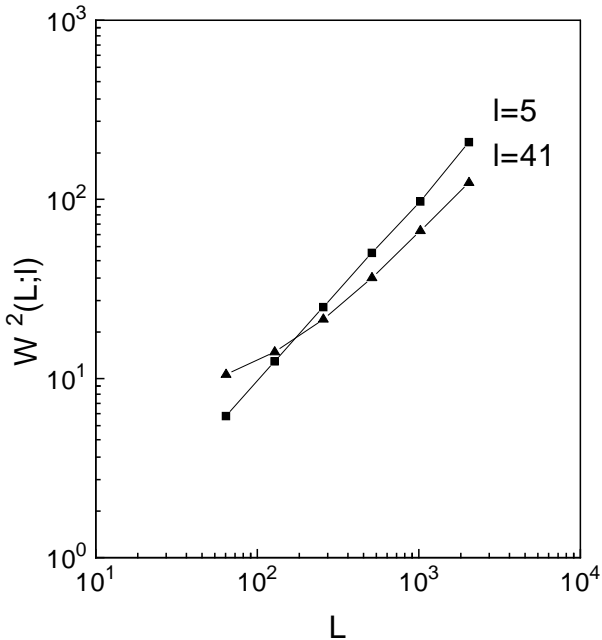}}
\caption{Double-logarithmic plot of $W^2(L;\ell)$ versus $L$ 
for typical subset sizes $\ell=5$ and $41$.} 
\label{fig6}
\end{figure}
\begin{figure}
\smallskip
\centerline{\epsfxsize=8cm \epsfbox{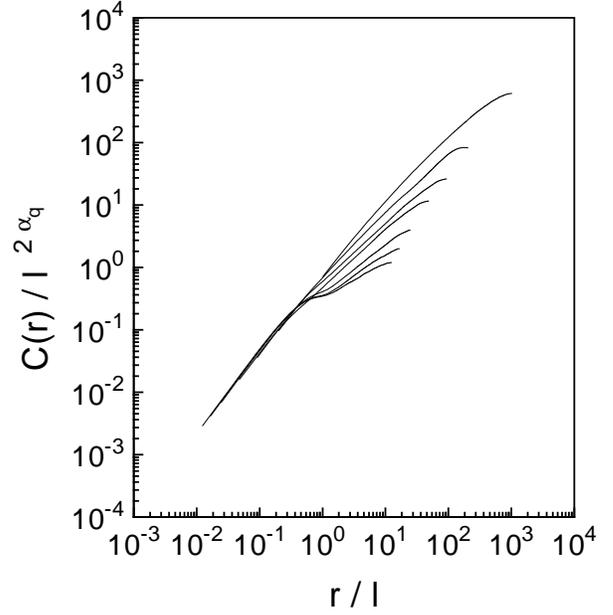}}
\caption{
Double-logarithmic plot of $C(r)/\ell^{2\alpha_q}$ versus
$r/\ell$ for subset sizes $\ell=5,11,21,41,61,81$ and 141. 
The data are for system size $L=2048$.} 
\label{fig7}
\end{figure}
\begin{figure}
\smallskip
\centerline{\epsfxsize=8cm \epsfbox{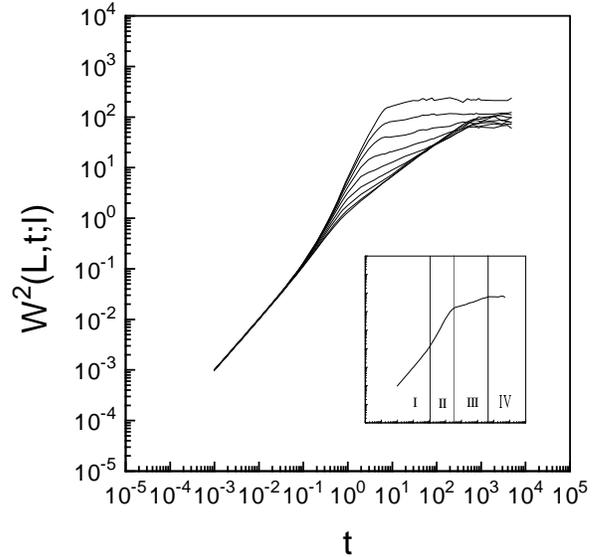}}
\caption{
Double-logarithmic plot of $W^2(L,t;\ell)$ versus time $t$
for system size $L=1024$. The data are for subset sizes 
$\ell=3,5,9,17,33,65,129,257$ and 513.
Inset: Distinct four regimes are observed for the case 
of $L=1024$ and $l=65$.} 
\label{fig8}
\end{figure}
\begin{figure}
\smallskip
\centerline{\epsfxsize=8cm \epsfbox{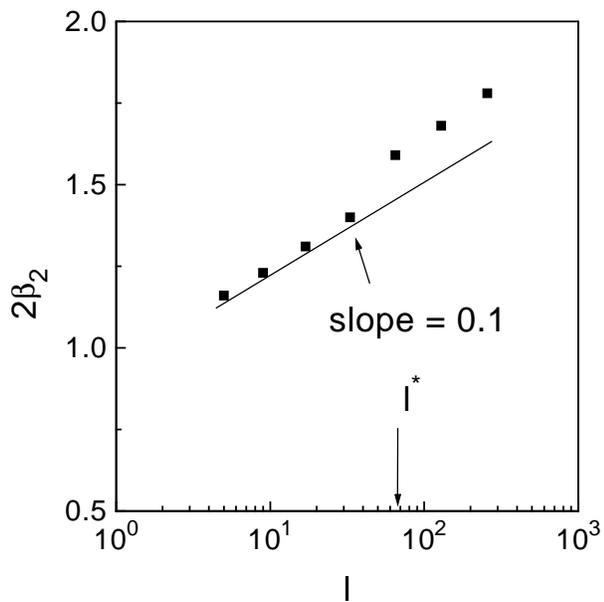}}
\caption{
Semi-logarithmic plot of $2\beta_2$ versus $\ell$ for $L=1024$. 
The data seem to be on a straight line with slope 0.1 up to the 
characteristic subset size $\ell^*$.} 
\label{fig9}
\end{figure}
\begin{figure}
\smallskip
\centerline{\epsfxsize=8cm \epsfbox{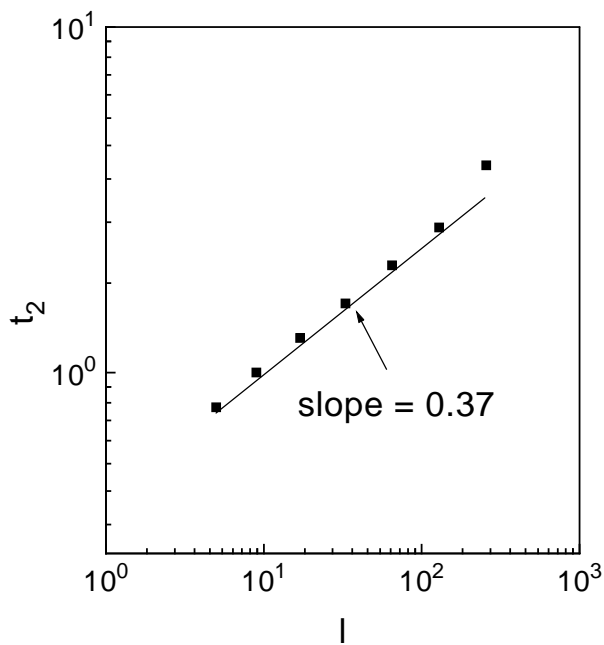}}
\caption{
Double-logarithmic plot of $t_2$ versus $\ell$ for $L=1024$. 
The data seem to be on a straight line with slope $0.37$.} 
\label{fig10}
\end{figure}
\begin{figure}
\smallskip
\centerline{\epsfxsize=8cm \epsfbox{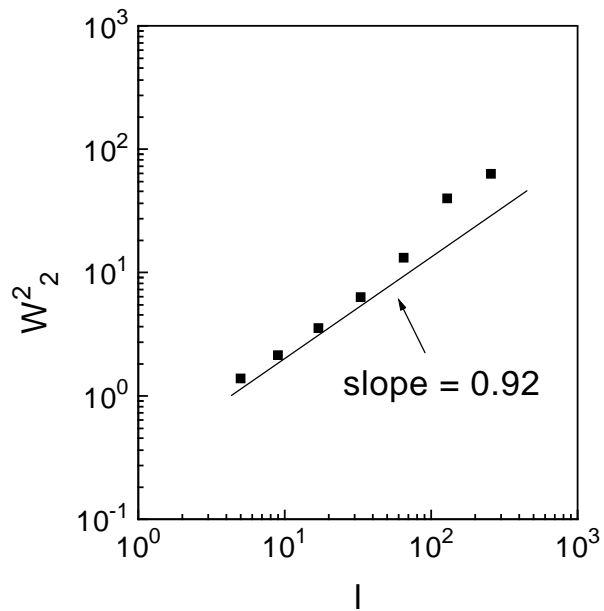}}
\caption{
Double-logarithmic plot of ${W_2}^2$ versus $\ell$ for $L=1024$. 
The data seem to be on a straight line with slope $0.92$
up to the characteristic subset size $\ell^*$.} 
\label{fig11}
\end{figure}
\begin{figure}
\smallskip
\centerline{\epsfxsize=8cm \epsfbox{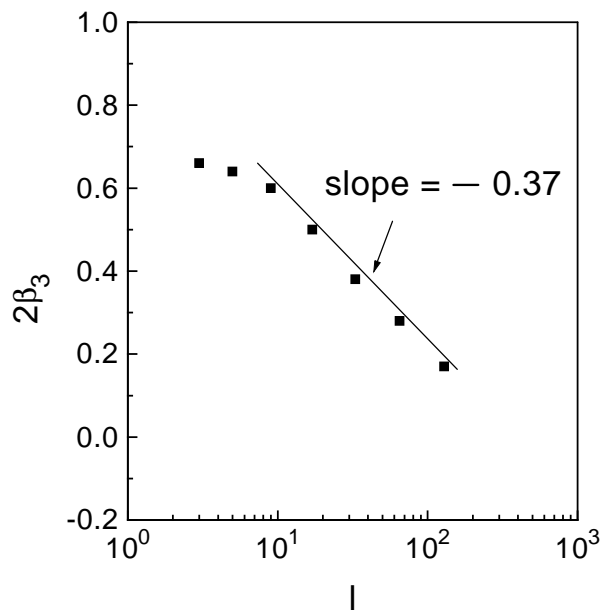}}
\caption{
Semi-logarithmic plot of $2\beta_3$ versus $\ell$ for $L=1024$. 
The line has the slope $-0.37$, and the data are for 
$\ell = 3, 5, 9 ,17, 33, 65$ and $129$, respectively.    
} 
\label{fig12}
\smallskip
\end{figure}
\begin{figure}
\centerline{\epsfxsize=8cm \epsfbox{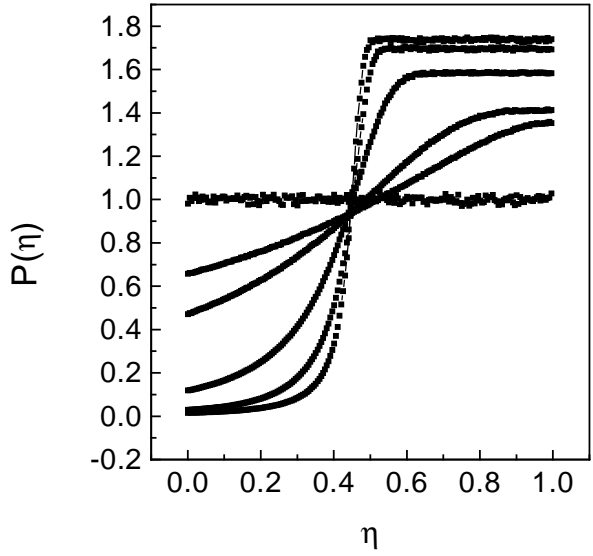}}
\caption{
Plot of the random number distribution after saturation
for $L=256$. The data are for $\ell = 1,3,5,27,129$ and $256$,
respectively. 
} 
\label{fig13}
\end{figure}

\end{multicols}
\end{document}